\documentstyle[preprint,prl,aps]{revtex} 

\newcommand{\Tr}{{\rm Tr}\,}
\newcommand{\CD}{{\cal D}}
\newcommand{\CN}{{\cal N}}

\newcommand{\CP}{{\cal P}}
\newcommand{\CF}{{\cal F}}
\newcommand{\CZ}{{\cal Z}}

\newcommand{\CS}{{\cal S}}
\newcommand{\CW}{{\cal W}}

\begin{document}
\draft
\title{\begin{flushright}
{\small\hfill AEI-063\\
\hfill hep-th/9804199}\\
\end{flushright}
Finite Yang-Mills Integrals}
\author{Werner Krauth \footnote{krauth@physique.ens.fr }}
\address{CNRS-Laboratoire de Physique Statistique,
Ecole Normale Sup\'{e}rieure\\
24, rue Lhomond\\ F-75231 Paris Cedex 05, France}
\author{Matthias Staudacher \footnote{matthias@aei-potsdam.mpg.de }
\footnote{Supported in part by EU Contract FMRX-CT96-0012} }
\address{
Albert-Einstein-Institut, Max-Planck-Institut f\"{u}r
Gravitationsphysik\\ Schlaatzweg 1\\  D-14473 Potsdam, Germany }
\maketitle
\begin{abstract}
We use Monte Carlo methods to directly evaluate $D$-dimensional
$SU(N)$ Yang-Mills partition functions reduced to zero Euclidean
dimensions, with and without supersymmetry.  In the non-supersymmetric
case, we find that the integrals exist for $D=3$, $N \geq
4$ and $D=4$, $N \geq 3$ and, lastly, $D \geq 5$, $N \geq 2$.  We
conclude that the $D=3$ and $D=4$ integrals exist in the large $N$
limit, and therefore lead to a well-defined, new type of Eguchi-Kawai
reduced gauge theory. For the supersymmetric case, we check, up to
$SU(5)$, recently proposed exact formulas for the $D=4$ and $D=6$
D-instanton integrals, including the explicit form of the normalization
factor needed to interpret the integrals as the bulk contribution
to the Witten index.

\end{abstract}
\vspace{5.5cm}
\newpage
\narrowtext

In a recent study \cite{kns}, we developed Monte Carlo methods
in order to deal with dimensionally reduced Yang-Mills theories.
We directly work with the gauge potentials, in contradistinction
to more conventional numerical studies of lattice gauge theories.
Our initial interest was in establishing reliable methods for 
the numerical calculation of the bulk contribution to the Witten index
in supersymmetric field theories. Some further results in this
direction are presented below. The methods are actually even more 
efficient if applied to the non-supersymmetric case, and allow us
to settle the question of existence of Yang-Mills theory dimensionally
reduced to zero dimensions in a surprising way.

Let us first consider maximally
reduced D-dimensional $SU(N)$ Euclidean Yang-Mills theory.
The formal functional integral for the partition function then
becomes an ordinary integral:
\begin{equation}
\CZ_{D,N} = \int \prod_{A=1}^{N^2-1} \prod_{\mu=1}^{D}  
\frac{d X_{\mu}^{A}}{\sqrt{2 \pi}}
\exp \bigg[  \frac{1}{2} \Tr
[X_\mu,X_\nu] [X_\mu,X_\nu] \bigg].
\label{bosint}
\end{equation}
Since there are directions in the integration space which are not
suppressed, these integrals were generally believed to be ill-defined.
Indeed, it is very easy to show that e.g.~for $D=2$ the integral diverges
for all $N$. Nevertheless, in \cite{kns} we obtained an
astonishing result: For gauge group $SU(2)$
the exact result reads
\begin{equation}
\CZ_{D,2}=
\left\{
\begin{array}{cc}
\infty & \;\;\;D \leq 4 \\
&\\
2^{-\frac{3}{4}D -1} \frac{\Gamma(\frac{D}{4}) \Gamma(\frac{D-2}{4})
\Gamma(\frac{D-4}{4})}{\Gamma(\frac{D}{2}) \Gamma(\frac{D-1}{2})
\Gamma(\frac{D-2}{2})} & \;\;\;D \geq 5
\end{array}  
\right.
\label{nosusy}
\end{equation}
Therefore, the reduced bosonic theory is {\it not necessarily}
divergent. It is natural to ask how eq.(\ref{nosusy}) generalizes
if $N \geq 3$. The methods used to derive 
eq.(\ref{nosusy}) are specific to $SU(2)$, and no analytic
result is known for higher gauge groups with $N \geq 3$, except
for $D=2$, as mentioned.
However, we can modify the methods of \cite{kns} to decide
the question of existence by Monte Carlo evaluation. Our
results suggest the following intriguing answer:
\begin{eqnarray}
\CZ_{D,N} < \infty \;\;\;\;\;\; &{\rm for}\;\;\;&
\left\{
\begin{array}{cc}
D=3 & \;N \geq 4 \\
&\\
D=4 & \;N \geq 3 \\
&\\
D \geq 5 & \;N \geq 2
\end{array}  
\right.
\cr
& & \cr
Z_{D,N} = \infty \;\;\;\;\;\; & {\rm otherwise} & 
\label{converge}
\end{eqnarray}
In particular, this suggests that a well defined large $N$ limit exists
in dimensions $D=3$ and, most interestingly, $D=4$. 
It opens the exciting possibility that
appropriate large $N$ correlation functions computed for 
the model (\ref{bosint})
can be related to large $N$ Yang-Mills field theory through the
Eguchi-Kawai mechanism \cite{ek}. E.g.~, one could consider Wilson loop
operators such as
\begin{equation}
\CW(L,T)=
\langle \frac{1}{N} \Tr e^{i L X_1} e^{i T X_D} e^{-i L X_1} e^{-i T X_D} 
\rangle
\label{wilson}
\end{equation}
in the limit $N \rightarrow \infty$. 
In fact, our model eq.(\ref{bosint})
is a new type of continuum (as opposed to lattice) Eguchi-Kawai model.
Reduced models are frequently plagued by the need to introduce
quenching, but our models are already in the weak
coupling phase, so this does not appear to be a problem.
However, these important questions are
beyond the scope of the present analysis.

Let us explain how to obtain convincing evidence for the result
eq.(\ref{converge}).
In \cite{kns} we computed {\em ratios} of integrals for different dimensions
$D$. Such a strategy is rendered necessary by the strong fluctuations
of the integrand in eq.(\ref{bosint}). One key point in our procedure
\cite{kns} 
consists in {\em compactifying} the integrals: after introducing
polar coordinates $(x_1, \ldots, x_d) = (\Omega_d, R)$ ($d= D (N^2-1)$ being 
the total dimension of the integral), we exactly perform the R-integration.
In the present  (bosonic)  case of eq(\ref{bosint}), $\CZ_{D,N}$ can 
be written as 
\begin{equation}
\CZ_{D,N} = \frac { \int {\cal D}\Omega_d~z_{D,N}(\Omega_d) }
{ \int {\cal D}\Omega_d  }
\end{equation}
with 
\begin{equation}
z_{D,N}(\Omega_d)=
2^{-(N^2-1) \frac{D}{2} -1}
\frac{ \Gamma\Big((N^2-1) \frac{D}{4}\Big) }
{ \Gamma\Big((N^2-1) \frac{D}{2}\Big) } \times
\frac{1}
{\Big[ \CS(\Omega_d,1) \Big]^{\frac{D}{4} (N^2-1)}}
\label{havelz2}
\end{equation}
To obtain the absolute value of $\CZ_{D,N}$, we now consider a
series of interpolating functions $z_{D,N}^{\alpha_i}(\Omega_d)$
with  $1=\alpha_0 <
\alpha_1 < \ldots <\alpha_{l-1} < \alpha_l=0$
(notice that $z_{D,N} \geq 0$). These interpolating functions allow
us to connect $z_{D,N}(\Omega_d)$ to a constant in much the same
way as we can (sometimes) compute the free energy of a statistical
physics system by simulating at various temperatures from $\infty$
down to the temperature of interest. The term with $\alpha_l=0$ 
plays the role of the exactly solvable high-temperature limit, since we can 
integrate the constant function $z_{D,N}^{\alpha_l} = 1$
analytically on the $d$-sphere.
We now write 
\begin{equation}
\CZ_{D,N}= 
\bigg[\frac{\int{\cal D}\Omega_d~z^{\alpha_1} 
\times z^{\alpha_0 - \alpha_1}}{\int{\cal D}\Omega_d~z^{\alpha_1}} \bigg]
\bigg[\frac{\int{\cal D}\Omega_d~z^{\alpha_2} 
\times z^{\alpha_1 - \alpha_2}}{\int{\cal D}\Omega_d~z^{\alpha_2}} \bigg]
\ldots
\bigg[\frac{\int{\cal D}\Omega_d~1 
\times z^{\alpha_{l-1} }}{\int{\cal D}\Omega_d~1} \bigg]
\label{alphahop} 
\end{equation} 
Each
of the terms $[\;]$ in eq(\ref{alphahop}) is computed in a separate
Monte Carlo run, in which points 
are sampled according to  $ \pi(\Omega) \sim z^{\alpha_i}$,
using the Metropolis algorithm. The fluctuations of the operator to be 
evaluated,  
$z^{\alpha_{i-1} - \alpha_i}$, are much damped
if $\alpha_{i-1} \sim \alpha_i$. 
In practice, we have used $l= 10$,
corresponding to the number of work stations available to us.
Judiciously chosen values of $\alpha_i$, closely spaced as we approach $1$,
allow us to compute any finite bosonic integral with ease. As a 
simple check, we have computed $\CZ_{N=2,D=5}$ ({\it cf} eq.(\ref{nosusy}))
to within $1 \%$ precision
in about $30$ minutes of individual CPU time on $10$ machines.

The finiteness of the integrals was also checked with the extremely
powerful {\em qualitative} Monte Carlo method described in \cite{kns}.
In that method, we solely monitor the autocorrelation function of the 
Metropolis random walk. The cases analytically known to diverge
were easily identified ($D=2$ for all $N$, $D=2,3,4$ for $N=2$). Of these
cases, the $N=2, D=4$ random walk appears less divergent, which 
agrees with the analytic result that the divergence is marginal 
(the $N=2, D=4+\epsilon$ integral exists). A similar behavior is 
observed for $N=3, D=3$, which we believe to be marginally divergent
as well.

Using the techniques just described, we are able to present
the following table:

\begin{center}
Table 1: Direct evaluation of bosonic Yang-Mills integrals\\
\vspace{0.5cm}
\begin{tabular}{|r||c|c|c||} \hline
$N$ & $D=2$     &  $D=3$     & $D=4$       \\ \hline
$2$ & $\infty$  &  $\infty$  & $ \infty$   \\
$3$ & $\infty$  &  $\infty$      & $1.9 \cdot 10^{-3}$        \\
$4$ & $\infty$  &  $6.9 \cdot 10^{-3}$      & $2.9 \cdot 10^{-8}$        \\
$5$ & $\infty$  &  $9.9 \cdot 10^{-7}$      & $2.1 \cdot 10^{-15}$        \\
$6$ & $\infty$  &  $4.8 \cdot 10^{-12}$     & $4.1 \cdot 10^{-25}$        \\ 
\hline
\end{tabular} 
\end{center}

The computation becomes simpler both with $D$ 
(it is completely trivial for $D>5$)
and $N$. We are thus confident about the statements expressed
for large $N$ and arrive at the prediction (\ref{converge}).
This is the central observation of the present paper.

Turning now to the supersymmetric case, the integrals reduced to zero
dimensions read 
\begin{equation}
\CZ_{D,N}^{\CN}:=\int \prod_{A=1}^{N^2-1} 
\Bigg( \prod_{\mu=1}^{D} \frac{d X_{\mu}^{A}}{\sqrt{2 \pi}} \Bigg) 
\Bigg( \prod_{\alpha=1}^{\CN} d\Psi_{\alpha}^{A} \Bigg)
\exp \bigg[  \frac{1}{2} \Tr 
[X_\mu,X_\nu] [X_\mu,X_\nu] + 
\Tr \Psi_{\alpha} [ \Gamma_{\alpha \beta}^{\mu} X_{\mu},\Psi_{\beta}]
\bigg].
\label{susyint}
\end{equation}
The only possible dimensions are $D=3,4,6,10$ corresponding to
$\CN=2,4,8,16$ real supersymmetries, respectively. Integrating out the
fermions, we find an integral that differs from eq.(\ref{bosint}) only
by a modified measure:
\begin{equation}
\CZ_{D,N}^{\CN} = \int \prod_{A=1}^{N^2-1} \prod_{\mu=1}^{D}  
\frac{d X_{\mu}^{A}}{\sqrt{2 \pi}}
\exp \bigg[  \frac{1}{2} \Tr
[X_\mu,X_\nu] [X_\mu,X_\nu] \bigg]~
\CP_{D,N}(X),
\label{int}
\end{equation}
The integrand is weighted by a very special 
homogeneous polynomial $\CP_{D,N}$ of degree 
$k=(D-2) (N^2-1)=\frac12 \CN (N^2-1)$ in the variables $X_{\mu}^A$.
See e.g.~\cite{kns} for further details.
In a very recent calculation, Moore et.al.~\cite{moore},
heavily using cohomological field theory techniques (and 
therefore supersymmetry), evaluated, for $D=4,6,10$ analytically 
a class of related integrals which are initially defined in Minkowski
space and are Wick rotated in the course of the evaluation.
These cohomologically deformed integrals 
are argued, after an appropriate 
prescription for the Wick rotation is found, to
take the same values as the integrals
(\ref{int}). They find the integrals to be of the form
\begin{equation}
\CZ_{D,N}^{\CN}=\CF_N \times \left\{ 
\begin{array}{ccc}
\frac{1}{N^2} & D=4, & \CN=4 \\ 
&\\
\frac{1}{N^2} & D=6, & \CN=8 \\
&\\ 
\sum_{m | N} \frac{1}{m^2} & D=10, & \CN=16
\end{array} \right.
\label{index}
\end{equation}
Let us point out that
the $D=10$ result had previously been conjectured
by Green and Gutperle \cite{greengut}.
For $SU(3)$, we had also performed a careful numerical check of the form
of eq.(\ref{index}) in \cite{kns}. 
$\CF_N$ is a group-theoretic factor not worked out explicitely
in \cite{moore}. In fact, it is unclear to us how to derive the 
the explicit form of $\CF_N$ from their approach.
In \cite{kns}, it was found to be
\begin{equation}
\CF_N=\frac
{2^{\frac{N(N+1)}{2}} \pi^{\frac{N-1}{2}}}
{2 \sqrt{N} \prod_{i=1}^{N-1} i!}.
\label{groupfactor}
\end{equation}

It is important to stress again our finding
that the Euclidean supersymmetric partition sums (\ref{susyint}),(\ref{int}) 
are not some formal mathematical entities, but perfectly converging ordinary
multiple integrals, just as their bosonic counterparts. It is very tempting
to speculate here once more 
that appropriate correlation functions computed for 
$N \rightarrow \infty$ could be related to correlators of the
full susy gauge field theories.

For completeness, let us sketch the derivation of (\ref{groupfactor}),
in particular since its precise relationship with the non-explicit 
normalization of \cite{moore} remains obscure.
In \cite{Sestern} is was argued how the $D$-dimensional {\it Euclidean}
matrix model emerges from the functional integral for the supersymmetric 
gauge quantum mechanics of $D-1$ matrices when computing
the bulk part of the Witten index
\begin{equation}
{\rm lim}_{\beta \rightarrow 0} \Tr (-1)^F e^{-\beta H}
\end{equation}
One needs to integrate over the group $SU(N)$ in order to
project onto gauge invariant
states. This integration becomes, in the limit $\beta \rightarrow 0$,
an integration over the hermitean generators of the group
\begin{equation}
\CD U \rightarrow \frac{1}{\CF_N} \prod_{A=1}^{N^2-1} 
\frac{d X^A_D}{\sqrt{2 \pi}},
\label{haarflat}
\end{equation}
where $\CF_N$ is precisely the constant relating the bulk part of the
index to the Euclidean matrix model. It is easily found if we 
investigate how the normalized Haar measure $\CD U$ and the flat measure
$ \prod_{A=1}^{N^2-1} d X^A_D$ act on class functions. The latter
depend only on the eigenvalues $z_i=e^{i \lambda_i}$ of the unimodular
unitary matrices $U$. One easily checks (e.g.~by verifying orthonormality
of group characters) that group integration on class functions $f(U)$
is performed
as
\begin{equation}
\int \CD U f(U) =
\frac{1}{N!} \oint \prod_{k=1}^{N} \frac{d z_k}{2 \pi i z_k}
\Delta(z_i) \Delta_(z_i^*) 2 \pi \delta_p(\lambda_1+ \ldots \lambda_N) 
f(z_1,\ldots,z_N)
\label{haar}
\end{equation}
where $\delta_p$ is the $2 \pi$-periodic $\delta$-function and
$\Delta(z_i)=\prod_{i<j}(z_i-z_j)$.
On the other hand, the flat measure for the hermitean matrices $X_D$ becomes
\begin{equation}
\prod_{A=1}^{N^2-1} \frac{d X^A_D}{\sqrt{2 \pi}}=
\frac{2^{\frac{N(N-1)}{2}}}{\prod_{i=1}^{N} i!} \sqrt{N}
\prod_{k=1}^{N} \frac{d \lambda_k}{\sqrt{\pi}} \Delta^2(\lambda_i)
\sqrt{\pi} \delta(\lambda_1+ \ldots + \lambda_N)
\label{flat}
\end{equation}
where the factors are easily checked by computing a Gaussian integral
normalized to one 
(the $\lambda_i$ are the eigenvalues of $X_D$).
Then, replacing (since $\beta \rightarrow 0$, see \cite{Sestern})
\begin{equation}
z_i=1+i \lambda_i+\ldots,
\end{equation}
comparing (\ref{haar}) and (\ref{flat})
and multiplying by an extra factor of $N$, due to the fact that
the unitary matrices localize on $N$ values because of invariance under
the center of the group, we can read off from eq.(\ref{haarflat})
the result eq.(\ref{groupfactor}) for $\CF_N$.

In order to test the group factor  (\ref{groupfactor}) as well as the 
form (\ref{index}) of the $D=4$ bulk part of the Witten index,
we have applied the methods of direct Monte Carlo evaluation explained
above to the $D=4$ and $N \leq 5$ integrals. 
As conjectured in \cite{kns}, we find that the integrand
is always positive if $D=4$. This means that we can use the method
of direct evaluation described earlier. 
The results are presented in table 2.

\begin{center}
Table 2: Direct evaluation of the $D=4$ D-instanton integral\\
\vspace{0.5cm}
\begin{tabular}{||r|c|c|c||} \hline
$N$ & eqs.(\ref{index}),(\ref{groupfactor})  &  Monte Carlo  & error \\ \hline
 2      & $ 1.25$   &  1.25  & $<$0.01        \\
 3      & $ 3.22$   &  3.22  & $<$0.01        \\
 4      & $ 7.42$   &  7.6  & $\pm$ 0.2        \\ 
 5      & $10.04$   & 10.2  & $\pm$ 0.2   \\ \hline
\end{tabular}
\end{center}

We have also checked that the $D=6$ integrals equal the $D=4$ integrals to
good precision by the ratio method of \cite{kns}.

Finally, let us mention the open problem of the evaluation of the
supersymmetric $D=3$ integral. In \cite{kns}, we conjectured
\begin{equation}
\CZ^{\CN=2}_{D=3,N}=0
\label{threedim}
\end{equation}
This result is trivial for $N$ even, but non-trivial for odd $N$.
We suspect that the supersymmetric $D=3$ integrals are absolutely
convergent (except for $SU(2)$, and possibly $SU(3)$), just like their
non-supersymmetric counterparts, {\it cf} eq.(\ref{converge}), and that
their well-defined value is given by eq.(\ref{threedim}) \cite{ks2}.
Another interesting
conjecture was presented in \cite{moore}, where it was suggested that
a modified $D=3$ integral $\tilde{\CZ}^{\CN=2}_{D=3,N}$, where the action
is complemented by a Chern-Simons term, leads to
\begin{equation}
\tilde{\CZ}^{\CN=2}_{D=3,N}=\CF_N~\frac{1}{N^2}.
\label{chern}
\end{equation}
Actually, for the solvable case of $SU(2)$, it can be shown that
eq.(\ref{chern}) is {\it not} true, since one easily finds
$\tilde{\CZ}^{\CN=2}_{D=3,N=2}=\infty$.
However, our bosonic result (\ref{converge}) suggests that the 
susy integrals modified by Chern-Simons exist for at least $N \geq 4$,
and therefore it is feasible to test the conjecture eq.(\ref{chern})
for generic gauge groups by our approach \cite{ks2}.

\acknowledgements
We thank I.K.~Kostov, S.~Shatashvili and especially
H.~Nicolai for useful discussions.
M.~S. thanks the LPS-ENS Paris for hospitality. 
This work
was supported in part by the EU under Contract FMRX-CT96-0012.

\end{document}